\documentstyle[12pt,epsfig]{article}
 1
 1
 1

\begin{document}
\bibliographystyle{unsrt}
%
\def\lta{\;\raisebox{-.5ex}{\rlap{$\sim$}} \raisebox{.5ex}{$<$}\;}
\def\gta{\;\raisebox{-.5ex}{\rlap{$\sim$}} \raisebox{.5ex}{$>$}\;}

%

\newcommand{\permille}{$^0 \!\!\!\: / \! _{00}\;$}
\newcommand{\GeV}{GeV}
 
\newcommand{\mt}{m_{t}}
\newcommand{\mtt}{m_{t}^2}
\newcommand{\mw}{M_{W}}
\newcommand{\mww}{M_{W}^{2}}
\newcommand{\md}{m_{d}}
\newcommand{\mb}{m_{b}}
\newcommand{\mbb}{m_{b}^2}
\newcommand{\mc}{m_{c}}
\newcommand{\mh}{m_{H}}
\newcommand{\mhh}{m_{H}^2}
\newcommand{\mz}{M_{Z}}
\newcommand{\mzz}{M_{Z}^{2}}

\newcommand{\lra}{\leftrightarrow}
 
\newcommand{\ie}{{\em i.e.}}
\def\Ww{{\mbox{\boldmath $W$}}}  
\def\B{{\mbox{\boldmath $B$}}}         
\def\nn{\noindent}

\newcommand{\sinsq}{\sin^2\theta}
\newcommand{\cossq}{\cos^2\theta}
\newcommand{\be}{\begin{equation}}
\newcommand{\ee}{\end{equation}}
\newcommand{\ba}{\begin{eqnarray}}
\newcommand{\ea}{\end{eqnarray}}
\newcommand{\eea}{\end{eqnarray}}

\newcommand{\nl}{\nonumber \\}
\newcommand{\eqn}[1]{Eq.(\ref{#1})}
\newcommand{\ibidem}{{\it ibidem\/},}
\newcommand{\into}{\;\;\to\;\;}
\newcommand{\wws}[2]{\langle #1 #2\rangle^{\star}}
\newcommand{\p}[1]{{\scriptstyle{\,(#1)}}}
\newcommand{\ru}[1]{\raisebox{-.2ex}{#1}}
\newcommand{\epem}{$e^{+} e^{-}\;$}
\newcommand{\tch}{$t\to c H\;$}
\newcommand{\tcz}{$t\to c Z\;$}
\newcommand{\tcg}{$t\to c g\;$}
\newcommand{\tchm}{$t\to c H$}
\newcommand{\tczm}{$t\to c Z$}
\newcommand{\tcgm}{$t\to c g$}
\newcommand{\tcfm}{$t\to c \gamma$}
\newcommand{\tcht}{t\to c H}
\newcommand{\tczt}{t\to c Z}
\newcommand{\tcgt}{t\to c g}
\newcommand{\tcft}{t\to c \gamma}

\newcommand{\tbwht}{t\to b W H}
\newcommand{\tbwzt}{t\to b W Z}
\newcommand{\tbwh}{$t\to b W H\;$}
\newcommand{\tbwz}{$t\to b W Z\;$}
\newcommand{\tbwhm}{$t\to b W H$}
\newcommand{\tbwzm}{$t\to b W Z$}
\newcommand{\Gt}{\Gamma(t\to b W)}

\rightline{LC-TH-1999-012}
\rightline{ROME1-1281/99}
\rightline{December 1999}

\begin{center}
{\Large \bf The $t\to c H \;$ decay width 
in the standard model.
}
\end{center}
\bigskip

\begin{center}
{\large 
B.~Mele~$^{a,b},$
$\;\,$S.~Petrarca~$^{b,a} \;\,$
and $\;\,$A.~Soddu~$^{a,c}$ 
} \\

\bigskip\noindent
$^a$ INFN, Sezione di Roma 1, Rome, Italy\\
\noindent
$^b$ Rome University ``La Sapienza", Rome, Italy\\
\noindent
$^c$ University of Virginia, Charlottesville, VA, USA

\end{center}
\bigskip
\begin{center}
{\bf Abstract} \\
\end{center}
{\small 
The \tch decay width has been computed in the standard
model with a light Higgs boson. 
The corresponding branching fraction 
has been found to be in the range $B(\tcht)\simeq 10^{-13} \div 10^{-14}$
for $\mz \lta \mh \lta 2\mw$. 
Our results correct the numerical evaluation usually quoted in the literature.
}
\bigskip

The one-loop flavor-changing transitions, \tcgm, \tcfm, \tcz and \tchm,
are particularly interesting, among the top quark rare decays.
Indeed, new physics, such as supersymmetry, 
an extended Higgs sector and heavier-fermion families
could conspicuosly affect the rates for this decays.
In the standard model (SM), these processes are in general quite 
suppressed due to the
Glashow-Iliopoulos-Maiani (GIM) mechanism, 
controlled by the light masses of
the $b, s, d$ quarks circulating in the loop. 
The corresponding branching fractions $B_i=\Gamma_i/\Gamma_{T}$ 
are further decreased 
by the large total decay width $\Gamma_{T}$ of the top quark.
The complete calculations of the one-loop
flavour-changing top decays have been performed,
before the top quark experimental observation,
in the paper by Eilam, Hewett and Soni ~\cite{sonid} 
(also based on Eilam, Haeri and Soni ~\cite{soniu}).
Assuming  $\mt=175$GeV, the value of the
total width $\Gamma_{T}\simeq \Gt$ is $\Gamma_{T} \simeq 1.55$ GeV,
and one gets from ref.~\cite{sonid} 
\be
B(\tcgt)\simeq 4\cdot 10^{-11}, \;\;\;\; \\
B(\tcft)\simeq 5\cdot 10^{-13}, \;\;\;\;  \\
B(\tczt)\simeq 1.3\cdot 10^{-13}.
\label{giust}
\ee
In the same ref.~\cite{sonid}, 
a much larger branching fraction  for
the decay \tch is presented as function 
 of the top and Higgs masses
(in  Fig. 1 the relevant Feynman graphs for this channel are shown).
\begin{figure*}[htbp]
\begin{center}
\vspace*{-4.cm}
\mbox{\epsfxsize=16cm\epsfysize=18.5cm\epsffile{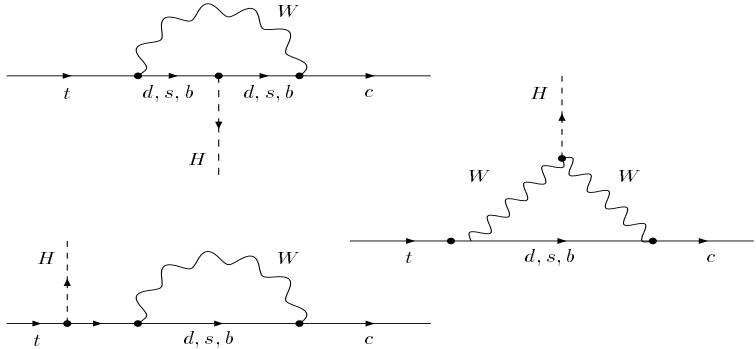}}
\vspace*{-8.5cm}
\caption{ Feynman graphs for the decay \tch in the unitary gauge
($\mc=0$ is assumed).
 }
\end{center}
\end{figure*}
For $\mt\simeq 175$ GeV and 40 GeV$\lta \mh \lta 2\mw$,
the value
\be
B(\tcht)\simeq 10^{-7} \div 10^{-8}
\label{sbagl}
\ee
is obtained, by means of the analytical formulae
presented in ref.~\cite{soniu}
for the fourth-generation quark decay  $b'\to bH$, in a 
theoretical framework assuming four flavour families.
Such  relatively large values for $B(\tcht)$ look surprising,
since the topology of the Feynman graphs  for the
 different one-loop channels is similar,
and  a GIM suppression, governed by the down-type
quark masses, is acting in all the decays.

In order to clarify the situation,
we recomputed from scratch the complete 
analytical decay width for \tch, as described in \cite{nostro}.
The corresponding numerical results for $B(\tcht)$, when
$\mt = 175$GeV and $\Gt \simeq 1.55$ GeV, 
are reported in Table 1. 
We used  $\mw=80.3$GeV, $\mb=5$GeV, $m_s=0.2$GeV,
and for the Kobayashi-Maskawa matrix elements
$|V_{tb}^*V_{cb}|=0.04$.
Furthermore, we assumed 
$|V_{ts}^*V_{cs}|=|V_{tb}^*V_{cb}|$. As a consequence,
the $\md$ dependence in the amplitude drops out.

\noindent
Our results are several orders of magnitude smaller 
than the ones reported in the literature.
In particular, for $\mh \simeq \mz$ we obtain
\be
B_{new}(\tcht) \simeq 1.2\cdot 10^{-13}  
\label{nostro}
\ee
to be compared with the corresponding value presented in ref.~\cite{sonid}
\be
B_{old}(\tcht) \simeq 6\cdot 10^{-8}.  
\label{loro}
\ee

In order to trace back the source of this inconsistency, we 
performed a thorough study of the analytical formula in eq.~(3)
of ref.~\cite{soniu}, for the decay width of the fourth-family
down-type quark $b'\to bH$, that is  the basis
for the numerical evaluation of $B(\tcht)$ presented in 
ref.~\cite{sonid}.
The result of this study was that we agreed with the
analytical computation in \cite{soniu},
but we disagreed with the numerical evaluation of $B(\tcht)$ in \cite{sonid}.

\noindent
The explanation for this situation 
can be ascribed to 
some error in the computer 
code used by the authors of ref.~\cite{sonid} to work out their Fig.~3. 
This explanation has been confirmed to us by one of the authors of 
ref.~\cite{sonid} (J.L.H.),
and by the erratum  appeared consequently \cite{erradue}, whose evaluation
we now completely agree with.

\begin{figure*}[htbp]
\begin{center}
\vspace*{-2.cm}
\mbox{\epsfxsize=16cm\epsfysize=18.5cm\epsffile{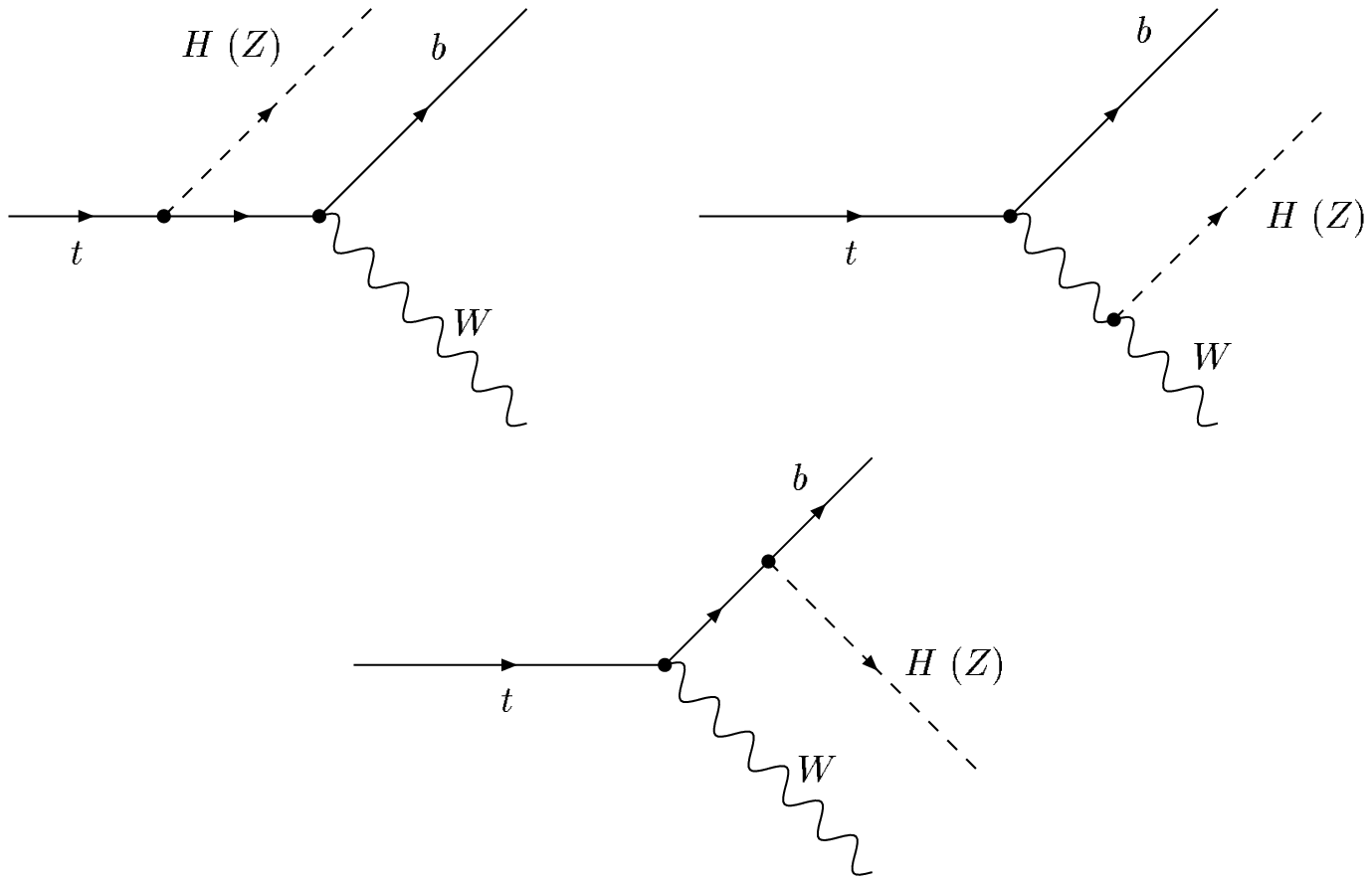}}
\vspace*{-9.5cm}
\caption{  Feynman graphs for the decay \tbwh (\tbwzm).
 }
\end{center}
\end{figure*}
 In the following we give some euristic considerations
 useful in order to understand the correct order of magnitude
 of the rate for the decay \tch. 
The comparison
between the rates for \tcz and \tch and the corresponding
rates for the tree-level decays \tbwz and \tbwhm, when $\mh \simeq \mz$
can give some hint on this order of magnitude.
In fact, the latter channels can be considered a sort of lower-order 
{\it parent}
processes for the one-loop decays, as can be seen
in Fig.~2, where the relevant Feynman graphs are shown. 
\begin{table}[tbhp]  
\begin{center}  
\begin{tabular}{|c|c|} \hline   
		 &                                           \\   
$ m_H\;(GeV) $ &  $ B(\tcht)  $ \\    
[2.mm]\hline\hline  
	&                                                 \\   
  80  &  $ 0.1532\cdot10^{-12} $ \\ [2.mm]\hline   
	&                                                 \\   
  90  &  $ 0.1169\cdot10^{-12} $ \\ [2.mm]\hline   
	&                                                 \\   
  100 &  $ 0.8777\cdot10^{-13} $ \\ [2.mm]\hline   
	&                                                \\   
  110 &  $ 0.6452\cdot10^{-13} $ \\ [2.mm]\hline   
	&                                                 \\   
  120 &  $ 0.4605\cdot10^{-13} $ \\ [2.mm]\hline   
	&                                                 \\   
  130 &  $ 0.3146\cdot10^{-13} $ \\ [2.mm]\hline   
	&                                                 \\   
  140 &  $ 0.1998\cdot10^{-13} $ \\ [2.mm]\hline   
	&                                                 \\   
  150 &  $ 0.1105\cdot10^{-13} $ \\ [2.mm]\hline   
	&                                                 \\   
  160 &  $ 0.4410\cdot10^{-14} $ \\    
	&                                                 \\   
  [2.mm]\hline\hline   
\end{tabular}   
\vspace*{1.4cm}
\caption{ Branching ratio for the decay 
\tch versus $\mh$. 
We assume $\mt=175$GeV and $m_c=1.5$GeV.
}
\end{center}  
\end{table}    
Indeed, the Feynman graphs for \tcz and \tch can be obtained 
by recombining the final $b$ quark and $W$ into a $c$ quark
in the three-body decays \tbwz and \tbwhm, respectively, 
and by adding analogous
contributions where the $b$ quark is replaced by the $s$ and $d$ quarks.
Then, the depletion of the
\tch rate with respect to the parent \tbwh rate
is expected to be of the same
order of magnitude of the depletion  of \tcz with respect to \tbwzm,
for $\mh \simeq \mz$. In fact, the GIM mechanism acts in a similar way
in the one-loop decays into $H$ and $Z$.

The \tbwz and \tbwh decay rates
have been computed, taking into account crucial
$W$ and $Z$ finite-width  effects, in ref.~\cite{realiu}.
 For $\mh \simeq \mz$, the two widths are 
comparable. In particular, for $\mt\simeq 175$GeV, one has \cite{realiu}
\be
B(\tbwzt) \simeq 6\cdot 10^{-7}   \; \; \; \;\; \; \; \;\;  
B(\tbwht) \simeq 3\cdot 10^{-7}.
\label{brtree}
\ee
From \cite{sonid}, $B(\tcht) \simeq 6\cdot 10^{-8}$ 
for $\mh \simeq \mz$.
Accordingly, the ratio of the one-loop and tree-level decay rates is
\be
r_H \equiv \frac{B(\tcht)}{B(\tbwht)} \sim 0.2  
\label{rhiggs}
\ee
to be confronted with 
\be
r_Z \equiv \frac{B(\tczt)}{B(\tbwzt)} \sim 2\cdot 10^{-7}.
\label{rzeta}
\ee
On the other hand,  $r_H$ 
and $r_Z$ are related to the quantity
\be
\left( \frac{g}{\sqrt{2}} |V_{tb}^*V_{cb}| \frac{m_b^2}{M_W^2} \right)^2   
\sim 10^{-8}   
\label{equa}
\ee 
(where $V_{ij}$ are the Kobayashi-Maskawa matrix elements)
arising from 
the higher-order in the weak coupling and the GIM suppression 
mechanism of the one-loop decay width. 
The large discrepancy between the value of the ratio $r_H$ 
in eq.~(\ref{rhiggs}) and what was expected from the factor in 
eq.~(\ref{equa}), which, on the other hand, is supported by the value of 
$r_Z$, was a further indication that the values for $B(\tcht)$
reported in eq.~(\ref{sbagl}) could be incorrect.

\noindent
Indeed, the new value of $B(\tcht)$ in eq.~(\ref{nostro}) 
gives  $r_H \sim 4 \cdot 10^{-7}$.

In conclusion,  we have  pointed out that one of the numerical 
results of ref.~\cite{sonid} establishing a relatively large branching 
ratio for the decay
\tch in the SM has been  overestimated.
The correct numerical estimates are shown in Table 1.
We find $B(\tcht)\simeq 1 \cdot 10^{-13}\div 4\cdot 10^{-15}$
for $\mz \lta \mh \lta 2\mw$.
Such a small rate will not be measurable even at the highest
luminosity accelerators that are presently conceivable.
An eventual experimental signal in the rare $t$ decays will definitely 
have to be ascribed to some new physics effect.

\vskip 0.7cm
\noindent
We thank V.A.~Ilyin for discussions and suggestions.
\vskip 0.7cm


\end{document}